\documentclass[11pt,a4paper]{article}
\pdfoutput=1
\usepackage{jcappub}
\usepackage{slashed}

\def\be{\begin{equation}}
\def\ee{\end{equation}}
\def\bea{\begin{eqnarray}}
\def\eea{\end{eqnarray}}

\begin{document}

\title{A quantum-classical duality and emergent space-time}

\author{Vitaly Vanchurin}

\emailAdd{vvanchur@d.umn.edu}

\date{\today}

\affiliation{Department of Physics, University of Minnesota, Duluth, Minnesota, 55812 \\
Duluth Institute for Advanced Study, Duluth, Minnesota, 55804}

\abstract{We consider the quantum partition function for a system of quantum spinors and then derive an equivalent (or dual) classical partition function for some scalar degrees of freedom. The coupling between scalars is non-trivial (e.g. a model on $2$-sphere configuration space), but the locality structure of the dual system is preserved, in contrast to the imaginary time formalism. We also show that the measure of integration in the classical partition function can be formally expressed through relativistic Green's functions which suggests a possible mechanism for the emergence of a classical space-time from anti-commutativity of quantum operators.}

\maketitle

\section{Introduction}

It is well known that a $D$-dimensional quantum field theory at finite temperature can be mapped to a $D$+1-dimensional classical (or statistical) field theory. This quantum-to-classical correspondence is due to a simple observation that at equilibrium quantum systems are described by a quantum partition function 
\be
{\cal Z}_{\text{q}}[\beta] = Tr\left[ \exp\left( \beta\hat{H}_{\text{q}} \right )  \right]  \label{eq:quantum_partition1}
\ee
where the Hamiltonian operator $ \hat{H}_{\text{q}} $ can be formally interpreted as a generator of evolution in imaginary time $\beta$ and the trace implies that the evolution must be periodic \cite{Bloch,Zinn-Justion}. As a result the $D$-dimensional  quantum partition function can be expressed as a $D$+1-dimensional classical partition function on a space with one extra dimension (see Appendix \ref{sec:Appendix} for details).

The imaginary time formalism proved to be very useful for analyzing quantum field theories at finite temperature \cite{KirznitsLinde, Weinberg, Jackiw}, but there are certain limitations that do not permit the quantum-classical duality to be applied to more general systems. First of all, the inverse temperature parameter $\beta$ on the quantum side does not correspond to an inverse temperature parameter on the classical side, but to a parameter which appears inside of a classical Hamiltonian (more precisely $\beta$ is the size of extra dimension). Secondly, the classical system usually has many more degrees of freedom then the corresponding quantum system. For example, if the quantum system has a finite number of localized degrees of freedom, then in the corresponding classical system each of these degrees of freedom is represented by many copies that are spread out in the extra dimension. This is, of course, not very useful if one wishes to study a possible emergence of quantum mechanics from classical systems with {\it local} hidden variables. And finally, the imaginary time formalism works well as a mapping from quantum to classical systems, but not in the opposite direction since the dual classical systems are very restrictive (e.g. at least one dimension must be periodic). 

In this paper we shall extend the quantum-classical correspondence 
\be
 {\cal Z}_{\text{q}}[ \beta] = {\cal Z}_{\text{c}}[ \beta]  \label{eq:duality}
\ee
to mappings which preserve the locality structure of both quantum and classical theories. More precisely, we shall study quantum systems whose partition function $ {\cal Z}_{\text{q}}[ \beta]$  can be expressed as a classical partition function $ {\cal Z}_{\text{c}}[ \beta]$ with the same number of local degrees of freedom. The price that we will have to pay is that the discrete quantum degrees of freedom (e.g. spinors) would be mapped to continuous classical degrees of freedom (e.g. scalars). In terms of field theories we shall describe how certain $D$-dimensional quantum systems can be mapped to $D$-dimensional classical systems with parameter $\beta$ playing the role of an inverse temperature in both quantum and classical systems. An extra dimension on the classical side of the duality will emerge, but it will be Lorentzian and non-periodic, in contrast to the imaginary time formalism. 

The paper is organized as follows. In the next section we define a quantum system and introduce the abstract-indices notations that will be used throughout the paper. In Sec. \ref{sec:commuting} we construct a classical dual of the system in the limit when all of the degrees of freedom commute. In Sec. \ref{sec:anticommuting} we show that the anti-commutativity of quantum operators gives rise to a non-trivial measure of integration in the classical partition function and in Sec. \ref{sec:space-time} we show that the measure can be formally expressed as a causal (retarded) propagator on an emergent space-time. In Sec. \ref{sec:condition} we derive a sufficient condition for the existence of a classical dual and in Sec. \ref{sec:measure} we provide an example of the dual with a separable measure of integration. In Sec. \ref{sec:conclusion} we summarize and discuss the main results.

\section{Quantum system}\label{sec:spinors}

Consider a quantum system of $N$ fermionic subsystems (which we call spinors) described by operators $\hat{\gamma}^{j}_a$, satisfying the following commutation relation
\be
[\hat{\gamma}^ j_a, \hat{\gamma}^ k_b ] = 0  \label{eq:commutation1}
\ee
if $a\neq b$ and anti-commutation relation
\be
 \{ \hat{\gamma}^ j_a, \hat{\gamma}^ k_a \} = 2 \delta^{ j k} \hat{I} \label{eq:commutation2} 
\ee
where $a,b \in \{1, ..., N\}$ and $ j, k \in \{1, ..., D\}$. The spinor operators are also assumed to be Hermitian (or Majorana)
 \be
 \hat{\gamma}^ j_a = \hat{\gamma}^{ j\dagger}_a,
 \ee
and to satisfy a tracelessness condition
\be
Tr\left [\left(  \hat{\gamma}^{ j_1}_{a_1} ... \hat{\gamma}^{ j_{d_1}}_{a_1} \right ) \left ( \hat{\gamma}^{ j_1}_{a_2} ... \hat{\gamma}^{ j_{d_2}}_{a_2} \right ) ...  \left (  \hat{\gamma}^{ j_1}_{a_n} ... \hat{\gamma}^{ j_{d_n}}_{a_n} \right )\right ]=0.\label{eq:tracelessess}
\ee
where  $1\le a_1 < ... < a_n \le N$ and $1 \le j_1 <  j_2 < ... <  j_{d_k} \le D$ for all $k$. For example, a collection of spinor operators for $D=2$ can be represented by tensor products of Pauli $\hat{X}$ and $\hat{Y}$ matrices,
\bea
 \hat{\gamma}^1_1 &=& \hat{X}\otimes \hat{I} \otimes ... \otimes \hat{I}\notag\\
 \hat{\gamma}^2_1 &=& \hat{Y}\otimes \hat{I} \otimes ... \otimes \hat{I}\notag\\
 \hat{\gamma}^1_2 &=& \hat{I}\otimes \hat{X} \otimes ... \otimes \hat{I}\notag\\
 \hat{\gamma}^2_2 &=& \hat{I}\otimes \hat{Y} \otimes ... \otimes \hat{I}\\
 &...&\notag\\
 \hat{\gamma}^1_N &=& \hat{I}\otimes \hat{I} \otimes ... \otimes \hat{X}\notag\\
 \hat{\gamma}^2_N &=& \hat{I}\otimes \hat{I} \otimes ... \otimes \hat{Y}.\notag
\eea
Similarly, the tensor products of euclidean Dirac matrices (which can be constructed from tensor products of all three Pauli matrices) can be used to represent the spinor operators for $D=4$. Although the dimensionality $D$ of the subsystems is kept arbitrary the two cases with $D=1$ and $D=3$ are of particular importance as the respective quantum systems will turn out to be dual to simple classical models on $S^0$ (in Sec. \ref{sec:commuting}) and on $S^2$ (in Sec. \ref{sec:measure}) configuration/target spaces. 

From the spinor operators we construct a Hamiltonian operator
\be
\hat{H}_{\text{q}} = \sum_{ j_1 ... j_N \in \{0,...,D\}}   H_{ j_1...  j_N} \, \hat{\gamma}^{ j_1}_1 ... \hat{\gamma}^{ j_N}_N .
 \label{eq:Hamiltonian1}
\ee
where $\hat{\gamma}^0_a\equiv \hat{I}$ and all of the components $H_{ j_1...  j_N}$ are real numbers. Note that the Hamiltonian is not only non-homogeneous and non-local, but also non-$k$-local as it may include terms with an arbitrary number of $\hat{\gamma}^k_a$'s operators. The corresponding quantum partition function can be expanded in powers of   inverse temperature
\bea
{\cal Z}_{\text{q}}[ \beta] &=& Tr\left[ \exp\left( \beta\hat{H}_{\text{q}}  \right )  \right] \notag\\
& = &\sum_{n=0}^\infty \frac{\beta^n}{n!} \; Tr\left [ \left (\sum_{ j_1 ... j_N \in \{0,...,D\}}   H_{ j_1...  j_N} \, \hat{\gamma}^{ j_1}_1 ... \hat{\gamma}^{ j_N}_N  \right )^n \right ]
\label{eq:partition_expansion}
\eea
and then each power of Hamiltonian operator into a sum over all combinations of terms from the Hamiltonian
\be
Tr\left [ \left (\sum_{ j_1 ... j_N \in \{0,...,D\}}   H_{ j_1...  j_N} \, \hat{\gamma}^{ j_1}_1 ... \hat{\gamma}^{ j_N}_N \right )^n \right ] = \sum_{A}  h_A \;Tr\left [ \hat{\Gamma}^A \right ]\label{eq:power_expansion}
\ee
where $h_A$'s represent products of $H_{ j_1...  j_N}$'s components and $\hat{\Gamma}^A$'s  the corresponding products of the spinor operators. By upper-case letters (i.e. $A,B, C, ...$) we denote abstract-indices which include information about a particular order of terms from the Hamiltonian \eqref{eq:Hamiltonian1}. Note, however, that two different products  $h_A$ and $h_B$ of the same collections of $H_{ j_1...  j_N}$ components  would be the equal, but the corresponding products of operators $\hat{\Gamma}^A$ and  $\hat{\Gamma}^B$  may differ by a sign due to anti-commutativity condition \eqref{eq:commutation2}. For example, if $A$ represents $\left ( H_3^2  \hat{\gamma}^3_2 \right ) \left (H_2^2  \hat{\gamma}^2_2 \right)$ and $B$ represents $\left (H_2^2  \hat{\gamma}^2_2 \right) \left ( H_3^2  \hat{\gamma}^3_2 \right ) $ , then $h_A = H_2^2 H_3^1 = h_B$, but $\hat{\Gamma}^A = \hat{\gamma}^3_2 \hat{\gamma}^2_2 = -  \hat{\gamma}^2_2\hat{\gamma}^3_2  = - \hat{\Gamma}^B$.

It is convenient to define an ordered product of operators $:\hat{\Gamma}^A:$ (not to confuse with normal ordering) so that $\hat{\gamma}^{ k}_{a}$ appears to the left of $\hat{\gamma}^{ j}_{b}$ if either $a<b$ or if $a=b$ and $k<j$ (as for example in \eqref{eq:tracelessess}). Then 
\be
 \sum_{A}  h_A Tr\left [ \hat{\Gamma}^A \right ]  =  \sum_{A}   \theta(\hat{\Gamma}^A) h_A Tr\left [ :\hat{\Gamma}^A:\right ]  \label{eq:power_expansion2}
\ee
where 
\be
\theta(\hat{\Gamma}^A) = \begin{cases} 1 \;\;\;\;\;\;\text{if} \;\;\;  :\hat{\Gamma}^A: =  \hat{\Gamma}^A \\- 1 \;\;\;\text{if} \;\;\;  :\hat{\Gamma}^A :=  - \hat{\Gamma}^A. \end{cases}
\ee
Furthermore, if we define $\sigma(A)$ as a set of all abstract-indices which are equivalent to $A$ up to different combinations of terms from the Hamiltonian, then 
\be
 \sum_{A}   \theta(\hat{\Gamma}^A) h_A Tr\left [ :\hat{\Gamma}^A:\right ]    =  \sum_{A}   \mu(A) h_A Tr\left [ :\hat{\Gamma}^A:\right ]  \label{eq:power_expansion3}
\ee
where 
\be
\mu(A) = \frac{1}{|\sigma(A)|} \sum_{B \in \sigma(A)} \theta(\hat{\Gamma}^B)
\label{eq:combinations}
\ee
and then \eqref{eq:power_expansion} can be rewritten as
\be
Tr\left [ \left (\sum_{ j_1 ... j_N \in \{0,...,D\}}   H_{ j_1...  j_N} \, \hat{\gamma}^{ j_1}_1 ... \hat{\gamma}^{ j_N}_N \right )^n \right ] =   \sum_{A}   \mu(A) h_A Tr\left [ :\hat{\Gamma}^A:\right ].   \label{eq:power_expansion3}
\ee
For example, if $A$ represents $\left ( H_2^2  \hat{\gamma}^2_2 \right ) \left (H_3^1  \hat{\gamma}^3_1 \right)$, then 
\bea
h_A &=&H_2^2 H_3^1\notag\\
\hat{\Gamma}^A &=&  \hat{\gamma}^2_2 \hat{\gamma}^3_1\notag\\
:\hat{\Gamma}^A: &=& \hat{\gamma}^3_1 \hat{\gamma}^2_2 \\
\theta(\hat{\Gamma}^A)&=&1\notag\\
 \mu(A)&=& \frac{1}{2} \left ( 1 +1 \right ) =1,\notag
\eea
but if $A$ represents  $\left ( H_3^2  \hat{\gamma}^3_2 \right ) \left (H_2^2  \hat{\gamma}^2_2 \right)$, then  
\bea
h_A&=&H_3^2 H_2^2\notag\\
\hat{\Gamma}^A&=&\hat{\gamma}^3_2 \hat{\gamma}^2_2\notag\\
:\hat{\Gamma}^A:&=&\hat{\gamma}^2_2 \hat{\gamma}^3_2 \\
\theta(\hat{\Gamma}^A)&=&-1\notag\\
\mu(A)&=&\frac{1}{2}(1-1)=0.\notag
\eea
In general,  $\mu(A)$ can be calculated using commutation \eqref{eq:commutation1} and anti-commutation \eqref{eq:commutation2} relations, but the analysis is greatly simplified when all operators either commute (see Sec. \ref{sec:commuting}) or anti-commute (see Sec. \ref{sec:anticommuting}).

\section{Commuting operators}\label{sec:commuting}

Consider a system of $N$ quantum spinors \eqref{eq:Hamiltonian1} with $D=1$ whose quantum partition function is given by
\be
{\cal Z}_{\text{q}}[ \beta] =\sum_{n=0}^\infty \frac{\beta^n}{n!} \; Tr\left [ \left (\sum_{ j_1 ... j_N \in \{0,1\}}   H_{ j_1...  j_N} \, \hat{\gamma}^{ j_1}_1 ... \hat{\gamma}^{ j_N}_N   \right )^n \right ]\label{eq:quantum_partition2}
\ee
where $\hat{\gamma}^0\equiv \hat{I}$, and a system of  $N$ classical scalars $x_a$ whose classical partition function is
\be
{\cal Z}_{\text{c}}[ \beta]  =  {\cal N} \int \left ( \prod_a dx_a \rho(x_a)  \right ) \sum_{n=0}^\infty \frac{\beta^n}{n!}  \left ( \sum_{ j_1 ... j_N \in \{0,1\}}   H_{ j_1...  j_N} \, x^{ j_1}_1 ... x^{ j_N}_N \right)^n\label{eq:classical_partition0}
\ee
where $x^1_a \equiv x_a$ and $x^0_a \equiv 1$. The classical system \eqref{eq:classical_partition0} is dual to the quantum system \eqref{eq:quantum_partition2} if and only if the two partition functions are equal \eqref{eq:duality}, but since the equality must be satisfied for all $\beta$, an equivalent condition is 
\be
{\cal N} \int \left ( \prod_a d x_a \rho(x_a) \right )  \left ( \sum_{ j_1 ... j_N \in \{0,1\}}   H_{ j_1...  j_N} \, x^{ j_1}_1 ... x^{ j_N}_N \right)^n   = Tr\left [ \left (\sum_{ j_1 ... j_N \in \{0,1\}}   H_{ j_1...  j_N} \, \hat{\gamma}^{ j_1}_1 ... \hat{\gamma}^{ j_N}_N   \right )^n \right ]\notag
\ee
or using the abstract-indices notation \eqref{eq:power_expansion3}
\be
{\cal N} \int \left ( \prod_a d x_a \rho(x_a) \right )  \sum_{A}  h_A \;X^A   =  \sum_{A}   \mu(A) h_A Tr\left [ :\hat{\Gamma}^A:\right ].    \label{eq:matching2}
\ee
where $X^A$ is the corresponding products of the scalars. Since all of the spinor operators $\hat{\gamma}_a^1$'s commute the products of spinors are such that $:\hat{\Gamma}^A: =\hat{\Gamma}^A$ and $\mu(A) = 1$ for all $A$. And then by matching individual terms we get 
\be
{\cal N} \int \left ( \prod_a d x_a \rho(x_a) \right ) X^A      = Tr\left [ :\hat{\Gamma}^A: \right ].  \label{eq:matching4}
\ee

The ordered product of operators $:\hat{\Gamma}^A:$ either contains an even number of $\hat{\gamma}^1_a$ operators for every $a$, or there is at least one $a$ for which there is an odd number of $\hat{\gamma}^1_a$'s. In the latter case equation \eqref{eq:tracelessess} implies that $\hat{\Gamma}^A$ is traceless and so does not contribute to the partition function, and in the former case equation \eqref{eq:commutation2} implies that $\hat{\Gamma}^A = \hat{I}$. If we take the trace of identity to be 
\be
{\cal N}\equiv Tr\left [\hat{I} \right ] = D^N, \label{eq:normalization}
\ee
then the measure of integration $\rho(x_a)$ should be such that all odd statistical moments vanish and all even statistical moment are the same, i.e
\be
\int  \; (x_a)^{n} \rho(x_a) dx_a  = \begin{cases} 1 \;\; \text{if} \;\; n \;\; \text{is even} \\
0 \;\; \text{if} \;\; n \;\; \text{is odd}.   \end{cases}
\ee
But this is can be easily achieved with
\be
\rho(x_a) = \frac{\delta(x_a-1)+\delta(x_a+1)}{2} \label{eq:two_deltas}
\ee
which corresponds to a classical partition function
\be
{\cal Z}_{\text{c}}[ \beta]  = {\cal N} \int \left ( \prod_a d x_a  \frac{\delta(x_a-1)+\delta(x_a+1)}{2}  \right )\exp\left(\beta \sum_{ j_1 ... j_N \in \{0,1\}}   H_{ j_1...  j_N} \, x^{ j_1}_1 ... x^{ j_N}_N \right)
\ee
or simply
\be
{\cal Z}_{\text{c}}[ \beta] =  {\cal N}  \sum_{x_1 ... x_N \in \{1,-1 \}}  \;  \exp\left(\beta \sum_{ j_1 ... j_N \in \{0,1\}}   H_{ j_1...  j_N} \, x^{ j_1}_1 ... x^{ j_N}_N \right).
\ee

Therefore we conclude that the partition function of a quantum system whose Hamiltonian is built out of only $\hat{\gamma}^1_a$ operators can always be expressed as a classical partition function with Hamiltonian, 
\be
H_{\text{c}} = \sum_{ j_1 ... j_N \in \{0,1\}}   H_{ j_1...  j_N} \, x^{ j_1}_1 ... x^{ j_N}_N 
\ee
where $x_a$ are the classical spinors which take values plus or minus one, or equivalently classical scalars on $S^0$ target space. This also suggests that the eigenvalues of the Hamiltonian $\hat{H}_{\text{q}}$ are simply related to its components $H_{ j_1  j_2 ...  j_N}$, i.e.
\be
E_x =   \sum_{ j_1 ... j_N \in \{0,1\}}  H_{ j_1  j_2 ...  j_N} \, x^{ j_1}_1 x^{ j_2}_2 ... x^{ j_N}_N,
\ee
where $x \in \{-1,1\}^N$.

\section{Anti-commuting operators}\label{sec:anticommuting}

Next consider a single quantum spinor, i.e. $N=1$ but with arbitrary $D\ge1$, whose quantum partition function is 
\be
{\cal Z}_{\text{q}}[\beta] =\sum_{n=0}^\infty \frac{\beta^n}{n!} \; Tr\left [ \left (\sum_{ j \in \{1,...,D\}}   H_{ j} \, \hat{\gamma}^{ j}  \right )^n \right ]\label{eq:quantum_partition3}
\ee
where $\hat{\gamma}^0\equiv \hat{I}$, and a classical system of  $D$ scalars whose classical partition function is
\be
{\cal Z}_{\text{c}}[ \beta]  =  {\cal N} \int d^D x \rho(x) \sum_{n=0}^\infty \frac{\beta^n}{n!}  \left ( \sum_{ j \in \{1,...,D\}}   H_{ j} \, x^{ j} \right)^n. \label{eq:classical_partition3}
\ee
where by upper indices $j$ we now denote different scalars $x^j$. To find $\rho(x)$ such that the duality condition \eqref{eq:duality} is satisfied, we must once again match individual terms in the expansions of partition functions, i.e.
\be
{\cal N} \int d^D x \rho(x)   \left ( \sum_{ j \in \{1,...,D\}}   H_{ j} \, x^{ j}  \right)^n   = Tr\left [ \left ( \sum_{ j \in \{1,...,D\}}   H_{ j} \, \hat{\gamma}^{ j}  \right )^n \right ]\notag
\ee
or using the abstract-indices notation \eqref{eq:power_expansion3}
\bea
{\cal N} \int d^D x \rho(x)  \sum_{A}  h_A \;X^A   &=&  \sum_{A}   \mu(A) h_A Tr\left [ :\hat{\Gamma}^A:\right ],  \notag \\  \label{eq:matching7}
{\cal N} \int d^D x \rho(x)  X^A  &=&  \mu(A) Tr\left [ :\hat{\Gamma}^A:\right ] 
\eea
where $X^A$ is a given product of scalars $x^k$'s and  $:\hat{\Gamma}^A:$ is the corresponding ordered product of operators $\hat{\gamma}^k$'s. But since all odd statistical moments must vanish (because the trace of the corresponding product of operators would vanish \eqref{eq:tracelessess}) we obtain an equivalent condition 
\bea
{\cal N} \int d^D x \rho(x)  \prod_ k (x^k)^{2 n_ k}   &=&  \mu(A) Tr\left [ \prod_ k (\hat{\gamma}^k)^{2 n_ k}  \right ] \notag\\
\int d^D x \rho(x)  \prod_ k (x^k)^{2 n_ k}   &=&  \mu(A) 
    \label{eq:matching6}
\eea
for some even integers $2 n_k$'s which represent the number of operators $\hat{\gamma}^k$'s (or of scalars $x^k$'s) in the product of operators $\hat{\Gamma}^A$ (or in the product of scalars $X^A$). And so to calculate $\mu(A)$ (and then $\rho(x_a)$) we must sum over all combinations of anti-commuting spinor operators \eqref{eq:combinations} which are conveniently described by different terms in the multinomial expansions.  

Consider the following two multinomials: a sum of commuting scalars raised to some even power
\bea
 \left ( x^1+x^2+...+x^{D} \right)^{2K} &=& \sum_{m_1+...+m_{D}=2K} \frac{(m_1+ ...+ m_{D})! }{(m_1)! ... (m_{D})!}  (x^1)^{m_1} (x^2)^{m_2} ... (x^{D})^{m_{D}}  \notag\\
&=& \sum_{m_1+...+m_{D}=2K}  \frac{(\sum_ k  m_ k )! }{\prod_k ( m_ k)!}  (x^1)^{m_1} (x^2)^{m_2} ... (x^{D})^{m_{D}}  
 \label{eq:expansion_numbers}
\eea
and a sum of anti-commuting operators raised to the same even power
\bea
\left (\hat{\gamma}^{1}+\hat{\gamma}^{2}+...+\hat{\gamma}^{{D}}\right)^{2K} &=&\left ((\hat{\gamma}^{1})^2+(\hat{\gamma}^{2})^2+...+(\hat{\gamma}^{{D}})^2\right)^{K}   \notag\\ 
 &=&\sum_{n_1+...+n_{D}=K} \frac{( n_1+...+ n_{D})! }{(n_1)!  ... (n_{D})!}  (\hat{\gamma}^{1})^{2n_1} (\hat{\gamma}^{2})^{2n_2} ... (\hat{\gamma}^{{D}})^{2n_{D}}.\notag\\ 
 &=&\sum_{n_1+...+n_{D}=K} \frac{(\sum_ k  n_ k )! }{\prod_k ( n_ k)!}   (\hat{\gamma}^{1})^{2n_1} (\hat{\gamma}^{2})^{2n_2} ... (\hat{\gamma}^{{D}})^{2n_{D}}.\label{eq:expansion_operators}
\eea
Separate terms in the expansion of operators \eqref{eq:expansion_operators} represent products of $\hat{\gamma}^{k}$'s applied in different orders  (or combinations $\sigma(A)$)  and we are interested in products of $2n_1$ of $\hat{\gamma}^{1}$'s, $2n_2$ of $\hat{\gamma}^{2}$'s,  etc (see Eq. \eqref{eq:matching6}) The total number of such products (or $|\sigma(A)|$) is given by a multinomial coefficient in the expansion of scalars \eqref{eq:expansion_numbers} with $m_k=2n_k$, i.e.
\be
|\sigma(A)| =  \frac{(\sum_ k 2 n_ k )! }{\prod_ k (2 n_ k)!},\label{eq:multinomial}
\ee
but not all of them come with the same sign when compared to an operator ordered as $(\hat{\gamma}^{1})^{2n_1} (\hat{\gamma}^{2})^{2n_2} ... (\hat{\gamma}^{{D}})^{2n_{D}}.$ And   according to \eqref{eq:combinations} the sum of these signs (or $\theta(B)$'s) over all combinations of operators (or $B \in \sigma(A)$)  is what determines $\mu(A)$,
\bea
\mu(A) &=& |\sigma(A)|^{-1} \sum_{B \in \sigma(A)} \theta\left (B\right) \\
&=&   \frac{\prod_ k (2 n_ k)!}{(\sum_ k 2 n_ k )! } \frac{(\sum_ k n_ k )! }{\prod_ k n_ k ! } 
\eea
as is evident from \eqref{eq:expansion_operators}. 

This can be seen directly if we integrate \eqref{eq:expansion_numbers} weighted by ${\cal N} \rho(x)$ and then equate it to the trace of  \eqref{eq:expansion_operators},
\bea
 {\cal N}  \int d^{D} x \rho(x)  \left ( x^{1}+x^2+...+x^{D} \right)^{2K}  &=& Tr \left [\left (\hat{\gamma}^{1}+\hat{\gamma}^{2}+...+\hat{\gamma}^{{D}}\right)^{2K} \right ]\notag \\
\sum_{m_1+...+m_{D}=2K}   \frac{(\sum_ k m_ k )! }{\prod_ k m_ k ! } \int d^{D}x \rho(x)   \prod_ k (x^k)^{m_ k}   &=& \sum_{n_1+...+n_{D}=K} \frac{(\sum_ k n_ k )! }{\prod_ k n_ k ! } \frac{1}{ {\cal N} } Tr\left[ \prod_ k (\hat{\gamma}^{ k})^{2n_ k}\right]\notag \\
\sum_{m_1+...+m_{D}=2K}   \frac{(\sum_ k m_ k )! }{\prod_ k m_ k ! }  \int d^{D} x \rho(x)    \prod_ k (x^k)^{m_ k}   &=& \sum_{n_1+...+n_{D}=K} \frac{(\sum_ k n_ k )! }{\prod_ k n_ k ! }.
 \eea
Since the correct measure of integration  $\rho(x)$ must be such that all odd moments vanish the above condition can be rewritten as
 \bea
 \sum_{2 n_1+...+2 n_D=2K}   \frac{(\sum_ k 2 n_ k )! }{\prod_ k (2 n_ k)!}  \int d^{D} x \rho(x)    \prod_ k (x^k) ^{2 n_ k}   &=& \sum_{n_1+...+n_D=K} \frac{(\sum_ k n_k )! }{\prod_ k n_k ! } \label{eq:measure_condiiton}
 \eea
 where on the left hand side we substituted  $m$'s for $n$'s using $m_k=2n_k$. Then, in order for individual terms on both sides of \eqref{eq:measure_condiiton} to be equal the measure $ \rho(x)$ should be such that 
\be
\int   d^{D} x\, \rho(x) \;  \prod_ k (x^k) ^{2 n_ k}  = \frac{\prod_ k (2 n_ k)!}{(\sum_ k 2 n_ k )! }  \frac{(\sum_ k n_ k )! }{\prod_ k n_ k ! }  = \mu(A) \label{eq:moments} 
\ee
where $A$ can represent an arbitrary product of terms with $2n_1$ of $\hat{\gamma}^{1}$'s, $2n_2$ of $\hat{\gamma}^{2}$'s,  etc.

The moments generating function of $\rho(x)$ can be obtained directly from \eqref{eq:moments},
\bea
M(p_1, ..., p_{D}) &=& \sum_{n_1, ...,n_{D}}  \left ( \int d^{D} x \rho(x)    \prod_ k (x^k) ^{2 n_k} \right )  \frac{ p_1^{2n_1}... p_{D}^{2n_{D}}}{\prod_ k (2 n_k)!} \notag \\ 
&=& \sum_{n_1, ...,n_{D}}   \frac{\prod_ k (2 n_ k)!}{(\sum_ k 2 n_ k )! }  \frac{(\sum_ k n_ k )! }{\prod_ k n_ k ! } \frac{ p_1^{2n_1}... p_{D}^{2n_{D}}}{\prod_ k (2 n_ k)!} \notag \\ 
&=& \sum_{n_1, ...,n_{D}} \frac{1}{(2 \sum_ k n_ k )! } \frac{(\sum_ k n_ k )! }{\prod_ k n_ k ! }    p_1^{2n_1}... p_{D}^{2n_{D}} \notag \\ 
  &=& \sum_K \frac{1}{(2K )!}\sum_{n_1+...+n_D=K}  \frac{K! }{\prod_ k n_ k ! }  p_1^{2n_1}... p_{D}^{2n_{D}} \notag \\ 
  &=& \sum_K \frac{1}{(2K )!}\left( p_1^{2}+ ...+ p_{D}^{2} \right)^K\notag \\ 
 &=&  \cosh\left(\sqrt{ p_1^2+...+p_{D}^2}\right)
\eea
and then the corresponding characteristic function is
\be
M(i p_1, ..., i p_D) =  \cos\left(\sqrt{p_1^2+...+p_{D}^2}\right) = \cos\left(\sqrt{\sum_ k p_ k^2} \right )
\ee
and its inverse Fourier transform gives us the desired measure of integration
\be
\rho(x) = \int \frac{d^{D} p}{(2\pi)^{D}} \cos\left(\sqrt{\sum_ k p_ k^2} \right )  \exp\left (i \sum_ k x^k p_ k \right ).\label{eq:measure}
\ee

\section{Emergent space-time}\label{sec:space-time}

The measure of integration \eqref{eq:measure} for $D=1$ can be easily calculated,
\bea
\rho(x) &=& \frac{1}{2} \int \frac{d p}{2\pi}  \cos\left(\sqrt{p^2} \right ) \exp(i x p) \notag\\
&=& \frac{1}{2} \int \frac{d p}{2\pi}  \cos\left(p \right) \exp(i x p)  \notag\\
&=& \frac{1}{2} \int \frac{d p}{2\pi}  \left ( \exp(i {p}+ix p) + \exp(-i p +ix p) \right ) \notag\\
 &=& \frac{1}{2}\left (\delta(x+1)+\delta(x-1)\right),
\eea
which is in agreement with \eqref{eq:two_deltas}. As we will see shortly this result is due to the fact that the time derivative of a retarded Green's function of 1+1-dimensional d'Alembert operator is given by a sum of two delta-functions propagating in opposite directions on an emergent space-time. It turns out that the same statement is true in higher dimensions, i.e. $D>1$ or $D+1>2$, but the form of the Green's functions (and of the corresponding measures $\rho(x)$) is of course different.

To evaluate the integral in \eqref{eq:measure} for arbitrary ${D}$ we note that 
\be
\varphi(x^\mu) =\varphi(\vec{x}, x^0) \equiv  \int \frac{d^{D}p}{(2\pi)^{D}}\; \cos\left (x^0 \sqrt{\sum_ k (p_ k)^2}  \right ) \; \exp\left ( i \sum_ k p_ k x^ k \right )  \label{eq:Propagator}
\ee
solves a ${D}$+1-dimensional wave equation,
\be
 \left( (\partial_0)^2 - \sum_k {(\partial_ k)^2 } \right )\varphi(x^\mu) = 0,
\ee
with initial conditions
\bea
\varphi(\vec{x}, 0) &=& \delta^{({D})}(\vec{x})\\
 \partial_0  \varphi(\vec{x}, 0) &=&0
\eea
where
\bea
\partial_0 & \equiv & \frac{\partial}{\partial x^0}\\
\partial_k & \equiv & \frac{\partial}{\partial x^k}.
\eea
Then the solution of the ${D}$+1-dimensional wave equation is given by 
\be
\varphi(x^\mu)   = \int d^{{D}} y \,\partial_0 G(\vec{x}, x^0; \vec{y}, 0) \delta^{({D})}(\vec{y}) =\partial_0 G(\vec{x}, x^0)
\ee
where 
\be
G(x^\mu; y^\mu) =G(x^\mu-y^\mu)=  G(\vec{x}-\vec{y}, x^0 - y^0)
\ee
is the retarded Green's function (or retarded propagator) of ${D}$+1-dimensional d'Alembert operator. 

The retarded propagator can be expressed as a ${D}$-dimensional integral
\be
G(x^\mu;y^\mu) =   \int \frac{d^{D} p}{(2\pi)^{D}}\; \frac{\sin\left ((x^0-y^0) \, \omega(p)\right )}{ \omega(p)}\; \exp\left (i \,\sum_k p_k (x^k-y^k) \right )  \label{eq:Propagator}
\ee
where
\be
\omega(p) \equiv \sqrt{\sum_ k (p_k)^2}
\ee
or as a more symmetric ${D}$+1-dimensional integral 
\be
G(x^\mu; y^\mu) =   \int \frac{d^{{D}+1} p}{(2\pi)^{{D}+1}}\; \frac{\exp\left (i \,\sum_\mu p_\mu (x^\mu-y^\mu) \right ) }{\sum_\mu p_\mu p^\mu}
\ee
with appropriately chosen contour of integration (i.e. both poles shifted downwards) and $p^0 \equiv -p_0$. Then the solution is
\be
\varphi(x^\mu) = \partial_0 G(\vec{x}, x^0) =   \int \frac{d^{D+1} p}{(2\pi)^{D+1}}\; \frac{ i p_0 }{\sum_\mu p_\mu p^\mu}\exp\left(i \,\sum_\mu p_\mu (x^\mu-y^\mu) \right )
\ee
and we can define an extended  (into emergent ``temporal'' direction $T$) classical partition function
\bea
{\cal Z}_{\text{c}}[ \beta, T]  &=&    {\cal N} \, \int d^{D} x  \,\varphi(\vec{x}, T)   \exp\left( \beta \sum_{ j \in \{1,...,D\}}   H_{ j} \, x^{ j}  \right )\notag\\
& =&    {\cal N}  \int d^{D} x \,\partial_0 G(\vec{x}, T)  \exp\left( \beta\sum_{ j \in \{1,...,D\}}   H_{ j} \, x^{ j} \right )\label{eq:partition3}
\eea
which satisfies the desired duality condition 
\be
{\cal Z}_{\text{c}}[ \beta, 1]  =  {\cal Z}_{\text{q}}[ \beta]
\ee
and also normalization conditions
\bea
{\cal Z}_{\text{c}}[ \beta, 0]  = {\cal Z}_{\text{c}}[ 0, T]  = {\cal N}.
\eea

\section{Existence of duality}\label{sec:condition}

In the previous sections we obtained the correct measure of integration of the classical partition function for either commuting (i.e. $D=1$) or anti-commuting (i.e. $N=1$) degrees of freedom. Next we shall consider a more general system \eqref{eq:Hamiltonian1} with both commuting and anti-commuting terms (i.e. $N>1$ and $D>1$) described by the quantum partition function
\be
{\cal Z}_{\text{q}}[ \beta] =\sum_{n=0}^\infty \frac{\beta^n}{n!} \; Tr\left [  \left (\sum_{ j_1 ... j_N \in \{0,...,D\}}   H_{ j_1...  j_N} \, \hat{\gamma}^{ j_1}_1 ... \hat{\gamma}^{ j_N}_N   \right )^n \right ]\label{eq:quantum_partition3}.
\ee
The corresponding classical partition function (for $N D$ classical scalars $x^k_a$) can be defined in a similar manner
\be
{\cal Z}_{\text{c}}[\beta]  =  {\cal N} \int \left (\prod_a d^{D}x_a \right ) \rho(x) \sum_{n=0}^\infty \frac{\beta^n}{n!} \; \left (  \sum_{ j_1 ... j_N \in \{0,1\}}   H_{ j_1...  j_N} \, x^{ j_1}_1 ... x^{ j_N}_N\right )^n  \label{eq:classical_partition2},
\ee
but it is no longer clear if (or when) the duality condition \eqref{eq:duality} could be satisfied for some measure of integration
\be
\rho(x)= \rho(x^1_1, ... x^{D}_1, ..., x^1_N, ... x^{D}_N).
\ee
Indeed, by matching separate terms in \eqref{eq:quantum_partition3} and \eqref{eq:classical_partition2} we obtain
\be
{\cal N} \int \left ( \prod_a d^{D}x_a  \right )  \rho(x) \left (  \sum_{ j_1 ... j_N \in \{0,...,D\}}   H_{ j_1...  j_N} \, x^{ j_1}_1 ... x^{ j_N}_N\right )^n   = Tr\left [\left (\sum_{ j_1 ... j_N \in \{0,...,D\}}   H_{ j_1...  j_N} \, \hat{\gamma}^{ j_1}_1 ... \hat{\gamma}^{ j_N}_N   \right )^n  \right ]\notag
\ee
or using the abstract-indices notation \eqref{eq:power_expansion3}
\be
{\cal N} \int \left ( \prod_a d^{D}  x_a  \right )  \rho(x) \sum_{A}  h_A \;X^A    =   \sum_{A} \mu(A) h_A \;Tr\left [ :\hat{\Gamma}^A: \right ].\label{eq:matching10}
\ee
This suggests that the duality condition \eqref{eq:duality} can be satisfied only if individual terms in \eqref{eq:matching10} are equal, 
\be
{\cal N} \int \left ( \prod_a d^{D}  x_a \right )  \rho(x)   X^A   = \mu(A) Tr\left [ :\hat{\Gamma}^A: \right ].
\ee
Whenever the product of spinor operators $\hat{\Gamma}^A$ contains an even number $2n_k^a$ of $\hat{\gamma}^k_a$ for all $a$ and $k$ the trace is
\be
Tr\left [ :\hat{\Gamma}^A: \right ]= {\cal N}
\ee
and thus the even statistical moments must be given by 
\bea
{\cal N} \int \left ( \prod_a d^{D}  x_a \right ) & \rho(x)&   X^A    \;\;\;\;\;\;\;\;\;\;\;\; = {\cal N} \mu(A) \notag\\
\int  \left( \prod_{a} d^{D}  x_a   \right) & \rho(x) & \prod_{a, k}  \left (x^k_a \right ) ^{2 n^a_ k}   = \mu(A)  \label{eq:even_moments}
\eea
and all odd statistical moments must vanish. However, since the measure is uniquely determined from the statistical moments and the statistical moments are to be determined from $\mu(A)$, the measure can only exist if $\mu(A)$ is uniquely determined by non-negative integers ${n^a_ k}$'s. This puts a restriction on the quantum system \eqref{eq:quantum_partition3}, whose classical dual is \eqref{eq:classical_partition2}, which can be expressed as the following condition:
 \be
 :\hat{\Gamma}^A:=:\hat{\Gamma}^B: \;\;\;\;\Rightarrow\;\;\;\;\;\;\; \mu(A)=\mu(B).\label{eq:condition}
 \ee
In other words, even if $A$ and $B$ are not in the same combination class, i.e. $\sigma(A) \neq \sigma(B)$, but the corresponding products of operators are the same, i.e. $ :\hat{\Gamma}^A:=:\hat{\Gamma}^B: $, the statistical moments must also be the same, i.e. $\mu(A)=\mu(B)$. Note that \eqref{eq:condition} is only a sufficient condition as there might still be other dual classical systems which are not described by \eqref{eq:classical_partition2} (see for example imaginary time formalism in Appendix \ref{sec:Appendix}).  

\section{Separable measure}\label{sec:measure}

In the previous section we argued that the quantum system \eqref{eq:quantum_partition3}  will have a classical dual \eqref{eq:classical_partition2} if the condition \eqref{eq:condition} is satisfied. In this section we are going to study a particular example of the quantum system for which the classical dual not only exists, but the integration measure is also separable, i.e.
\be
\rho(x)= \rho(x^1_1, ... x^{D}_1, ..., x^1_N, ... x^{D}_N) = \prod_a \rho\left (x^1_a, ...,x^{D}_a\right ).\label{eq:measure_factorization}
\ee
Consider a Hamiltonian operator \eqref{eq:Hamiltonian1} with components which can be expressed as
\be
H_{ j_1...  j_N}  = \sum_{ k_1 ... k_N \in \{0,1\}} {\cal H}_{k_1...  k_N} \eta_{1, j_1}^{k_1} ...  \eta_{N, j_N}^{k_N}  \label{eq:coefficients}
\ee
where we assume that
\bea
\eta_{a, j}^{0} &=& \delta_{j0}\\
\eta_{a, 0}^{k} &=& \delta_{0k}
\eea
and,  without loss of generality, $ H_{0...0}=0$. Then the Hamiltonian operator can be rewritten as
\bea
\hat{H}_{\text{q}} &=& \sum_{ j_1 ... j_N \in \{0,...,D\}}   H_{ j_1...  j_N} \, \hat{\gamma}^{ j_1}_1 ... \hat{\gamma}^{ j_N}_N \\
&=& \sum_{ k_1 ... k_N \in \{0,1\}} {\cal H}_{k_1...  k_N} \left ( \sum_{ j_1 \in \{0,...,D\}} \eta_{1, j_1}^{k_1} \hat{\gamma}^{ j_1 }_1 \right ) ... \left ( \sum_{ j_N \in \{0,...,D\}} \eta_{N, j_N}^{k_N} \hat{\gamma}^{ j_N }_N \right )  \\
&=& \sum_{ k_1 ... k_N \in \{0,1\}} {\cal H}_{k_1...  k_N} \hat{\eta}^{k_1}_1 ... \hat{\eta}^{k_N}_N 
 \label{eq:Hamiltonian3}
\eea
where the combined operators $\hat{\eta}_a$'s are defined as linear combinations of spin operators
\be
\hat{\eta}_a = \hat{\eta}^{1}_a =  \sum_{ j \in \{0,...,D\}} \eta_{a, j}^1 \hat{\gamma}^{j}_a=  \sum_{ j \in \{1,...,D\}} \eta_{a, j}^1 \hat{\gamma}^{j}_a 
\ee
and 
\be
\hat{\eta}^{0}_a = \sum_{ j \in \{0,...,D\}} \eta_{a, j}^0 \hat{\gamma}^{j}_a = \sum_{ j \in \{0,...,D\}} \delta_{0j} \hat{\gamma}^{j}_a = \hat{\gamma}^{0}_a = \hat{I}.
\ee

Since the combined operators satisfy a commutation relation 
\be
\left [ \hat{\eta}_a, \hat{\eta}_b \right ] =0\label{eq:combined} 
\ee
we can essentially follow the analysis of Sec. \ref{sec:commuting} with classical partition function
\bea
{\cal Z}_{\text{c}}[\beta] &=&  {\cal N} \int \left (\prod_a d^{D}x_a  \rho(x_a) \right ) \sum_{n=0}^\infty \frac{\beta^n}{n!} \; \left (  \sum_{ j_1 ... j_N \in \{0,...,D\}}   H_{ j_1...  j_N} \, x^{ j_1}_1 ... x^{ j_N}_N\right )^n  \notag\\
&=&  {\cal N} \int \left (\prod_a d^{D}x_a  \rho(x_a) \right ) \sum_{n=0}^\infty \frac{\beta^n}{n!} \; \left ( \sum_{ k_1 ... k_N \in \{0,1\}} {\cal H}_{k_1...  k_N} \chi^{k_1}_1 ... \chi^{k_N}_N  \right )^n  \label{eq:classical_partition10},
\eea
where
\bea
\chi_a = \chi^{1}_a &=&  \sum_{ j \in \{1,...,D\}} \eta_{a, j}^1 x^{j}_a  \\
\chi^{0}_a &=&1.
\eea
Indeed, the duality condition \eqref{eq:duality} implies 
\be
{\cal N} \int\left (\prod_a d^{D}x_a  \rho(x_a) \right )  \left ( \sum_{ k_1 ... k_N \in \{0,1\}} {\cal H}_{k_1...  k_N} \chi^{k_1}_1 ... \chi^{k_N}_N \right )^n  = Tr\left [\left ( \sum_{ k_1 ... k_N \in \{0,1\}} {\cal H}_{k_1...  k_N} \hat{\eta}^{k_1}_1 ... \hat{\eta}^{k_N}_N \right)^n \right ]\ \notag
\ee
which can be satisfied (due to commutativity of the combined operators \eqref{eq:combined}) only if 
\be
{\cal N} \int d^{D}x_a  \rho(x_a)  \left ( \chi_a \right )^n  = Tr\left [\left (  \hat{\eta}_a \right)^n \right ]
\ee
or, equivalently, if
\be
{\cal N} \int d^{D}x_a  \rho(x_a)  \left ( \sum_{ j \in \{1,...,D\}} \eta^{1}_{j} x^{ j}_a \right )^n  = Tr\left [\left (   \sum_{ j \in \{1,...,D\}} \eta^{1}_{j} \hat{\gamma}^{ j}_a \right)^n \right ]\label{eq:matching9}
\ee
for all $a$ and $n$. But since all of $\hat{\gamma}^{ j}_a$ operators in \eqref{eq:matching9} anti-commute the corresponding measure is the same as in Sec. \ref{sec:anticommuting} which was shown to be given by  a retarded Green's function in Sec. \ref{sec:space-time}. Therefore, for a quantum system whose Hamiltonian components can be expressed as in \eqref{eq:coefficients}, the dual classical system is described by a classical partition function
\be
{\cal Z}_{\text{c}}[ \beta, T]  =    {\cal N}  \int \left (\prod_a  d^{D} x_a \,\partial_0 G(\vec{x}_a, T_a)   \right ) \exp\left( \beta \sum_{ j_1 ... j_N \in \{0,1\}}   H_{ j_1...  j_N} \, x^{ j_1}_1 ... x^{ j_N}_N\right ),\label{eq:partition6}
\ee
where $G(\vec{x}_a, T_a)$ is the retarded Green's function of ${D}$+1-dimensional d'Alembert operator (see Sec. \ref{sec:space-time}).  Note that the measure of integration is already normalized,
\be
\int \left (\prod_a  d^{D} x_a \,\partial_0 G(\vec{x}_a, T_a)   \right )   =1,
\ee
but it can be interpreted as a probability density only if it takes non-negative values. For example, when  $D=1$ 
\be
{\cal Z}_{\text{c}}[ \beta, T]   =    {\cal N} \int  \prod_a \left ( \frac{d x_a  }{2} \left( \delta(x^1_a-T_a)+\delta(x^1_a+T_a) \right)\right ) \exp\left( \beta H_{\text{c}} \right ),\label{eq:partition4}
\ee
in agreement with Sec \ref{sec:commuting}, or when $D=3$ 
\bea
{\cal Z}_{\text{c}}[ \beta, T]   =    {\cal N} \int \prod_a \left ( \frac{d^{3} x_a  }{4\pi T_a^2}  \delta\left(\sum_k \left(x^k_a\right)^2-T_a^2\right) \right ) \exp\left( \beta H_{\text{c}} \right ) \label{eq:partition5}.
\eea
Of course there is no reason to expect that the measure will remain positive for more general quantum systems and then the dual system defined in the similar manner would not be classical {\it per se} as the corresponding partition function would involve integration over configurations with negative probabilities.

\section{Discussion}\label{sec:conclusion}

In this paper we took a step towards extending the quantum-classical duality \eqref{eq:duality} beyond the imaginary time formalism. In particular we showed that the quantum partition function for a system of quantum spinors \eqref{eq:Hamiltonian1} with non-homogeneous, non-local, and non-$k$-local interactions can be described as a classical partition for some scalar degrees of freedom. The measure of integration in the classical partition functions is non-trivial, but the locality structure of the dual theories was preserved, in contrast to the imaginary time formalism. We derived a general sufficient condition for the existence of the duality \eqref{eq:condition} and gave three examples of the duality for quantum systems with only commuting (see Sec. \ref{sec:commuting}), with only anti-commuting  (see Sec. \ref{sec:anticommuting}), and with both commuting and anti-commuting (see Sec. \ref{sec:measure}) degrees of freedom. 

An interesting byproduct of our analysis was the realization that the non-trivial measure of integration in the classical partition function can be described using relativistic Green's functions (Sec. \ref{sec:space-time}). This suggest a possible (and quite general) mechanism for the emergence of a classical space-time from anti-commutativity of quantum operators which deserves a separate study (see, however, Refs. \cite{VanRaamsdonk:2010pw, Carroll, Noorbala, graph_flow} for other recent attempts to derive space-time from quantum mechanics). But since the duality mapping supposedly works both ways (from quantum to classical and from classical to quantum) the very same result can be interpreted as a possible mechanism for the emergence of anti-commutativity of quantum operators from Lorentzian symmetry of a classical space-time. And then it would be interesting to see if the phenomena can be responsible for the emergence of quantum mechanics from classical/statistical mechanics (see, for example, Refs. \cite{Adler, Caticha, tHooft, Entropic_Mechanics} for some recent attempts to derive quantum mechanics). 

The stumbling block for any classical theories with local hidden variables are the Bell's inequalities \cite{Bell1, Bell2}. It is hard to see how quantum effects, such as entanglement between qubits in an EPR pair, can be described using classical hidden variables that are also local \cite{Bell}. However, we have already seen that the quantum-to-classical mapping between equilibrium systems can preserve locality, and so it would be important to see if the locality can also be preserved away from the equilibrium. In  Sec. \ref{sec:space-time} we derived an extended partition function \eqref{eq:partition3} with a temporal parameter $T$ which describes the dynamics in an emergent space-time. For the equilibrium classical partition function the parameter $T$ had to be set to one, but the physical meaning of other values of $T$ remains unclear. Could it be that the extended partition function describes a non-equilibrium dynamics of the system from some zero entropy state at $T=0$ towards some maximum entropy state at $T=1$? And if so does this evolution follow the principle of the stationary entropy production that was recently proposed in \cite{Entropic_Mechanics}? We leave these questions for future work.

{\it Acknowledgments.} The work was supported in part by Foundational Questions Institute (FQXi).

\newpage
\appendix

\section{Imaginary time formalism}\label{sec:Appendix}

Consider a quantum system with a preferred tensor product factorization of Hilbert space into $N$ factors. The quantum partition function for such system can be written as
\be
{\cal Z}_{\text{q}}[\beta] = Tr\left[ \exp\left(\beta \hat{H}_{\text{q}} \right ) \right] = \sum_{j}  \langle j |\exp\left (\beta \hat{H}_{\text{q}} \right ) |j\rangle \label{eq:quantum_partition}
\ee
where the sum is taken over some set of orthonormal basis vectors 
\be
|j\rangle \equiv \bigotimes_{a=1}^N |j_a\rangle 
\ee
and
\be
j_a \in \{1,...,D\}.
\ee
This partition function can be evaluated as a path integral over imaginary time with periodic boundary conditions. By splitting the imaginary time $\beta$ into $T$ intervals we obtain
\be
{\cal Z}_{\text{q}}[\beta] = \sum_{j(1)}  \sum_{j(2)} ...  \sum_{j(T)} \langle j(T) |  e^{ \frac{\beta}{T} \hat{H}_{\text{q}}}  |j(1) \rangle \langle j(1)| e^{ \frac{\beta}{T} \hat{H}_{\text{q}}}  |j(2) \rangle... \langle x(T-1)| e^{ \frac{\beta}{T} \hat{H}_{\text{q}}}  |j(T) \rangle 
\ee
But this may be also interpreted as the classical partition function with summation taken over all configurations $j(1), j(2), .... , j(T)$ (or discretized paths) weighted by 
\be
p\left (j(1), j(2)... j(T)\right ) = \langle j(T) |  e^{ \frac{\beta}{T} \hat{H}_{\text{q}}}  |j(1) \rangle \langle j(1)| e^{ \frac{\beta}{T} \hat{H}_{\text{q}}}  |j(2) \rangle... \langle j(T-1)| e^{ \frac{\beta}{T} \hat{H}_{\text{q}}}  |j(T) \rangle.\label{eq:product}
\ee
Then what we have is two systems one quantum, described by partition function \eqref{eq:quantum_partition}, and another one classical (or more precisely statistical), described by 
\bea
{\cal Z}_{\text{c}}[\beta] &=&  \sum_{j(1)}  \sum_{j(2)} ...  \sum_{j(T)} p\left (j(1), j(2)... j(T)\right ) \notag\\
 &=&  \sum_{j(1)}  \sum_{j(2)} ...  \sum_{j(T)} \exp\left ({H}_{\text{c}}(\beta, j)  \right ) ,\label{eq:classical_partition}
\eea
where
\be
{H}_{\text{c}}(\beta, j) = \sum_{t=1}^T \log\left ( \left \langle j(t) \left |  \exp \left ( \frac{\beta}{T} \hat{H}_{\text{q}} \right) \right | j(t+1) \right \rangle \right) \label{eq:Hamiltonian}
\ee
such that the duality condition \eqref{eq:duality} is satisfied.

For sufficiently large $T$ the exponential can be approximated by a linear term, 
\be
\exp\left( \frac{\beta}{T} \hat{H}_{\text{q}} \right ) \approx \hat{I} +  \frac{\beta}{T} \hat{H}_{\text{q}}
\ee
and then 
\be
 \left \langle i \left |   \exp\left( \frac{\beta}{T} \hat{H}_{\text{q}}\right )  \right | j \right \rangle  \approx  \delta_{ij} + \frac{\beta}{T}  \langle i | \hat{H}_{\text{q}}  | j \rangle. \label{eq:transfer_matrix}
\ee
Equation \eqref{eq:transfer_matrix} describes what is known as transfer matrix, but if we want the product of elements of these matrices \eqref{eq:product} to represent probabilities or, equivalently, the classical Hamiltonian \eqref{eq:Hamiltonian} to be real,  certain restrictions must apply to its from. In particular, what we want is to choose basis vectors $|i \rangle$ so that the off-diagonal terms of the Hamiltonian $\langle i | \hat{H}_{\text{q}}  | j \rangle $ are non-negative and then to choose $T$ large enough so that the diagonal terms of the transfer matrix are also non-negative.


\newpage



\begin{thebibliography}{10}

\bibitem{Bloch}
Bloch, F. (1932),
 ``Zur Theorie des Austauschproblems und der Remanenzerscheinung der Ferromagnetika,'' 
 Z. Phys. 74 (5-6): 295-335.

\bibitem{Zinn-Justion}
Jean Zinn-Justin (2002),
`` Quantum Field Theory and Critical Phenomena,''
Oxford University Press

\bibitem{KirznitsLinde}
D.A. Kirznits and A.D. Linde,
``Macroscopic Consequences of the Weinberg Model,''
Phys. Lett. B42 (1972) 471

\bibitem{Weinberg}
 Weinberg, S. (1974),
 ``Gauge and Global Symmetries at High Temperature,'' 
 Phys. Rev. D. 9 (12): 3357-3378. 
 
 \bibitem{Jackiw}
L. Dolan, and R. Jackiw (1974), 
``Symmetry behavior at finite temperature,'' 
Phys. Rev. D. 9 (12): 3320-3341.



  
\bibitem{VanRaamsdonk:2010pw} 
  M.~Van Raamsdonk,
  ``Building up spacetime with quantum entanglement,''
  Gen.\ Rel.\ Grav.\  {\bf 42}, 2323 (2010)
 

\bibitem{Carroll}
  C.~Cao, S.~M.~Carroll and S.~Michalakis,
  ``Space from Hilbert Space: Recovering Geometry from Bulk Entanglement,''
Phys. Rev. D 95, 024031 (2017)

 \bibitem{graph_flow}
  V.~Vanchurin,
  ``Information Graph Flow: a geometric approximation of quantum and statistical systems,''
	Found.Phys. 48 (2018) no.6, 636-653
 
\bibitem{Noorbala}
M.~Noorbala,
``SpaceTime from Hilbert Space: Decompositions of Hilbert Space as Instances of Time, ''
Fortsch.Phys. 66 (2018) no.2, 1800002 

\bibitem{Adler} S. Adler, ``Quantum Theory as an Emergent Phenomenon'' (Cambridge UP, Cambridge, 2004)

\bibitem{Caticha} A. Caticha, ``Entropic Inference and the Foundations of Physics'' (EBEB 2012, S\~ao Paulo, Brazil)

\bibitem{tHooft} G. 't Hooft, ``The Cellular Automaton Interpretation of Quantum Mechanics'' (Springer, 2016)

\bibitem{Entropic_Mechanics}
  V.~Vanchurin,
  ``Entropic Mechanics,''
  arXiv:1901.07369 [cond-mat.stat-mech].


\bibitem{Bell1} S.J. Freedman; J.F. Clauser (1972). ``Experimental test of local hidden-variable theories''. Phys. Rev. Lett. 28 (938): 938-941

\bibitem{Bell2} Alain Aspect; Philippe Grangier; Gérard Roger (1981). ``Experimental Tests of Realistic Local Theories via Bell's Theorem'' Phys. Rev. Lett. 47 (7): 460-3


\bibitem{Bell} Bell, John (1964). ``On the Einstein Podolsky Rosen Paradox''. Physics. 1 (3): 195-200


\end{thebibliography}
\end{document}